\documentclass[12pt]{article}

\usepackage{geometry}
\usepackage{graphicx}
\usepackage[utf8]{inputenc}
\usepackage{amsmath,amscd}    % need for subequations
\usepackage{graphicx}   % need for figures
\usepackage{verbatim}   % useful for program listings
\usepackage{color}      % use if color is used in text
\usepackage{subfigure}  % use for side-by-side figures
\usepackage{hyperref}   % use for hypertext links, including those to external documents and URLs
\usepackage{bbm}
\usepackage{amsfonts}
\usepackage{mathtools}
\usepackage{amssymb}
\usepackage{tensor}
\begin{document}

%\maketitle
\author{
   Adam B. Gillard\\
   %\small\texttt{adam.gillard@yahoo.com}\\
  \and 
  Niels G. Gresnigt\footnote{Corresponding author: niels.gresnigt@xjtlu.edu.cn \newline \indent \;\;\;Department of Mathematical Sciences, Xi'an Jiaotong-Liverpool University, 111 Ren'ai Road, \indent\;\;\;Suzhou HET, Jiangsu, China 215123.}
}

\title{Three fermion generations with two unbroken gauge symmetries from the complex sedenions}
\maketitle

\begin{abstract}

We show that three generations of leptons and quarks with unbroken Standard Model gauge symmetry $SU(3)_c\times U(1)_{em}$ can be described using the algebra of complexified sedenions $\mathbb{C}\otimes\mathbb{S}$. A primitive idempotent is constructed by selecting a special direction, and the action of this projector on the basis of $\mathbb{C}\otimes\mathbb{S}$ can be used to uniquely split the algebra into three complex octonion subalgebras $\mathbb{C}\otimes \mathbb{O}$.  These subalgebras all share a common quaternionic subalgebra. The left adjoint actions of the 8 $\mathbb{C}$-dimensional $\mathbb{C}\otimes \mathbb{O}$ subalgebras on themselves generates three copies of the Clifford algebra $C\ell(6)$. It was previously shown that the minimal left ideals of $C\ell(6)$ describe a single generation of fermions with unbroken $SU(3)_c\times U(1)_{em}$ gauge symmetry. Extending this construction from $\mathbb{C}\otimes\mathbb{O}$ to $\mathbb{C}\otimes\mathbb{S}$ naturally leads to a description of exactly three generations.

\end{abstract}
\newpage
%\tableofcontents
\section{Introduction}

The role of division algebras, in particular the non-associative octonion algebra, in particle physics has a long history. Theorems by Hurwitz \cite{hurwitz1922komposition} and Zorn \cite{zorn1931theorie} guarantee that there are only four normed division algebras, which are also the only four alternative division algebras; the reals $\mathbb{R}$, complex numbers $\mathbb{C}$, quaternions $\mathbb{H}$, and the octonions $\mathbb{O}$. The first results relating the octonions to the symmetries of (one generation of) quarks goes back to the 1970s \cite{gunaydin1973quark,gunaydin1974quark}. The hypothesis that octonions might play a role in the description of quark symmetries follows from the observation that $\mathrm{Aut}(\mathbb{O})=G_2$ contains the physically important subgroup $SU(3)$, corresponding to the subgroup that holds one of the octonionic imaginary units constant. Octonions also describe geometry in 10 dimensions \cite{baez2002octonions}, which have made them useful in supergravity and superstring theories \cite{duff1998octonionic,baez2009division,baez2011division,anastasiou2014super}. Recently there has been a revived interest in using the division algebras to attempt to construct a theoretical basis for the observed Standard Model (SM) gauge groups and observed particle spectrum \cite{dixon2013division,furey2016standard,gresnigt2018braids,gresnigt2018braided,dray2010octonionic,burdik2017revisiting,burdik2018hurwitz,catto2018quantum}.

A Witt decomposition of the adjoint algebra of left actions $(\mathbb{C}\otimes\mathbb{O})_L\cong C\ell(6)$ of $\mathbb{C}\otimes\mathbb{O}$, generated from all composed left actions of the complex octonions on themselves decomposes $(\mathbb{C}\otimes\mathbb{O})_L$ into minimal left ideals \cite{ablamowicz1995construction}. The basis states of these ideals were recently shown to transform as a single generation of leptons and quarks under the unbroken unitary symmetries $SU(3)_c$ and $U(1)_{em}$ \cite{furey2016standard}\footnote{It is worth mentioning that a similar approach, using $C\ell(6)$ as a starting point, was concurrently developed by Stoica \cite{Stoica2018}.}. The main result presented here is that we instead consider a decomposition of  
$(\mathbb{C}\otimes\mathbb{S})_L$, where the sedenions $\mathbb{S}$ are the next Cayley-Dickson algebra after the octonions, which naturally extends these earlier results from one single generation to exactly three. 

Given that a considerable amount of the SM structure for a single generation of fermions can be realised in terms of the complex octonions $\mathbb{C}\otimes\mathbb{O}$, and its associated adjoint algebra of left actions $(\mathbb{C}\otimes\mathbb{O})_L\cong C\ell(6)$, a natural question is how to extend these encouraging results from a single generation to exactly three generations. Although it is possible to represent three generations of fermion states inside a single copy of $C\ell(6)$ \cite{furey2014generations,furey2018three}, in that case the physical states are no longer the basis states of minimal left ideals as for the case of a single generation. A three generation representation in terms of $C\ell(6)
$ therefore comes at the cost of giving up the elegant construction of minimal left ideals giving rise to a single generation. Instead, one might therefore look for a larger mathematical structure to describe three generations. 

Since describing a single generation requires one copy of the octonions, it seems reasonable to expect a three generation model to require three copies of the octonions. This makes the exceptional Jordan algebra $J_3(\mathbb{O})$ a natural candidate. This intriguing mathematical structure and its role in particle physics has recently been explored by various authors \cite{dubois2016exceptional,todorov2018deducing,dubois2019exceptional}.

A different approach is taken here. $\mathbb{C}$ and $\mathbb{O}$ are both division algebras, and starting from $\mathbb{R}$, all the other division algebras can be generated via the Cayley-Dickson process. Given that this process continues beyond the octonions, one may wonder if the next algebra in the sequence, the sedenions $\mathbb{S}$, may be a suitable mathematical structure to represent three generations of fermions. This paper confirms that the answer is yes. 

Starting with $\mathbb{C}\otimes\mathbb{S}$ we repeat the initial stages of the construction of \cite{furey2016standard}. We define a projection operator by choosing a special direction, and use it to obtain a split basis of 14 nilpotent ladder operators, defining two maximal totally isotropic subspaces (MTIS). But instead of using these nilpotents to construct two minimal left ideals of $(\mathbb{C}\otimes\mathbb{S})_L$, we instead observe that these 14 nilpotents uniquely divide into three sets of ladder operators, in such a way that each set provides a basis for a $\mathbb{C}\otimes\mathbb{O}$ subalgebra of $\mathbb{C}\otimes\mathbb{S}$. Our approach therefore gives rise to three copies of the octonions from the sedenions. In this construction one raising operator and one lowering operator is common to all three sets. The shared ladder operators generate a common quaternionic subalgebra $\mathbb{H}\cong C\ell(2)$. From the three sets of ladder operators we construct six ideals. These ideals transform as three generations of fermions and antifermions under the unbroken gauge symmetry $SU(3)_c\times U(1)_{em}$.  

The next section reviews the construction of one generation of fermions with unbroken $SU(3)_c\times U(1)_{em}$ gauge symmetry starting from the complex octonions $\mathbb{C}\otimes\mathbb{O}$. Then in Section \ref{Hideals} it is shown that the same construction can be used to describe three generations of leptons in terms of three $C\ell(2)$ subalgebras of $C\ell(6)$, generated from three $\mathbb{C}\otimes\mathbb{H}$ subalgebras of $\mathbb{C}\otimes\mathbb{O}$. The main results of the present paper are presented in Section \ref{3generations}, where three generations of fermions with unbroken $SU(3)_c\times U(1)_{em}$ gauge symmetries are constructed from the complex sedenions $\mathbb{C}\otimes\mathbb{S}$. The generators for these symmetries are constructed in Section \ref{symmetries}. 

%%%%%%%%%%%%%%%%%%%%%%%%%%%%%%%%%%%%%%%%%%%%%%%%%%%%%%%%%%%%%%%%%%%%%%%%%%%%%%%%%%%%%%%
\section{One generation of fermions as the basis states of the minimal left ideals of $(\mathbb{C}\otimes\mathbb{O})_L\cong C\ell(6)$}\label{octonions}

One of the main observations in  \cite{furey2016standard}\footnote{This work is also consistent with the works of Dixon \cite{dixon2013division,dixon2010division,dixon1990derivation,dixon2004division}. Those works include a copy of the quaternion algebra, recovering the full SM gauge group from the algebra $\mathbb{T}=\mathbb{R}\otimes\mathbb{C}\otimes\mathbb{H}\otimes\mathbb{O}$. The quaternionic factor in $\mathbb{T}$ is responsible for the broken $SU(2)$ chiral symmetry.} is that a single generation of fermions with two unbroken gauge symmetries $SU(3)_c$ and $U(1)_{em}$ can be represented starting from the complex octonions $\mathbb{C}\otimes\mathbb{O}$. The adjoint algebra of left actions $(\mathbb{C}\otimes\mathbb{O})_L$ of $\mathbb{C}\otimes\mathbb{O}$ on itself is the Clifford algebra $C\ell(6)$. In this section the minimal left ideals of $(\mathbb{C}\otimes\mathbb{O})_L\cong C\ell(6)$ are constructed using the Witt decomposition for $C\ell(6)$. The basis states of these ideals are then shown to transform as a single generation of leptons and quarks with unbroken $SU(3)_c$ and $U(1)_{em}$. More details of can be found in sections 4.5 and 6.6 of \cite{furey2016standard}.

\begin{figure}
\centering
\includegraphics[width=0.33\linewidth]{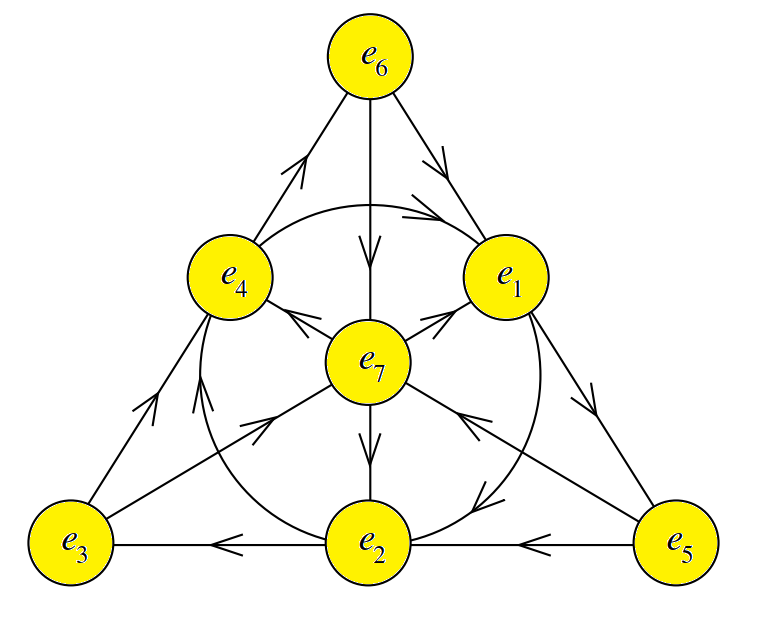}
\caption{Octonion multiplication represented using a Fano plane.}
\label{octonion}
\end{figure}

%%%%%%%%%%%%%%%%%%%%%%%%%%%%%%%%%%%%%%%%%%%%%%
The octonions $\mathbb{O}$ are spanned by the identity $1=e_0$ and seven anti-commuting  square roots of minus one $e_i$ satisfying
\begin{eqnarray}
e_ie_j=-\delta_{ij}e_0+\epsilon_{ijk}e_k,
\end{eqnarray}
where 
\begin{eqnarray}
e_ie_0=e_0e_i=e_i,\;e_0^2=e_0,
\end{eqnarray}
and $\epsilon_{ijk}$ is a completely antisymmetric tensor with value +1 when $ijk = 124,\;156$,$\;137$, $\;235$, $267,\;346,\;457 $. The multiplication of octonions is shown in Figure \ref{octonion}. The multiplication structure here follows that of \cite{furey2016standard}, but is not unique. In particular, the Fano plane in the earlier work \cite{gunaydin1973quark} is based on a different multiplication structure.  
Each pair of distinct points lies on a unique line of three points that are cyclically ordered. If $e_i$, $e_j$ and $e_k$ are cyclically ordered in this way then
\begin{eqnarray}
e_ie_j=e_k,\qquad e_je_i=-e_k.
\end{eqnarray}
Every straight line in the Fano plane of the octonions (taken together with the identity) generates a copy of the quaternions, for example $\left\lbrace 1,e_4,e_6,e_3\right\rbrace $. Furthermore $\left\lbrace 1,e_1,e_2,e_4\right\rbrace $ also gives a copy of the quaternions, making for a total of seven copies of the quaternions embedded within the octonions.

%%%%%%%%%%%%%%%%%%%%%%%%%%%%%%%%%%%%%%%%%%%%%%
One starts by choosing a privileged imaginary unit from the standard octonion basis ${e_0,e_1,...,e_7}$ and uses this to resolve the identity into two primitive idempotents. The common choice is $e_7$, so that
\begin{eqnarray}
1=\rho_+ + \rho_-=\frac{1}{2}(1+ie_7)+\frac{1}{2}(1-ie_7).
\end{eqnarray}
Multiplying the standard octonionic basis elements on the left by $\rho_+$ defines a split basis of nilpotents, and allows one to define the new basis vectors\footnote{following the convention of \cite{furey2016standard}.}
\begin{eqnarray}
\alpha_1\equiv\frac{1}{2}(-e_5+ie_4),\qquad \alpha_2\equiv\frac{1}{2}(-e_3+ie_1),\qquad \alpha_3\equiv \frac{1}{2}(-e_6+ie_2),
\end{eqnarray}
along with their hermitian conjugates 
\begin{eqnarray}
\alpha_1^{\dagger}\equiv\frac{1}{2}(e_5+ie_4),\qquad \alpha_2^{\dagger}\equiv\frac{1}{2}(e_3+ie_1),\qquad \alpha_3^{\dagger}\equiv \frac{1}{2}(e_6+ie_2).
\end{eqnarray}
The hermitian conjugate simultaneously maps $i\mapsto -i$ and $e_i\mapsto -e_i$. The new basis vectors span the two MTIS satisfying\footnote{The split basis in \cite{furey2016standard} differs slightly from the one given in \cite{gunaydin1973quark}. The split basis of the latter uses the complex conjugate $*$ instead of the hermitian conjugate $\dagger$, and satisfies the additional multiplication rules $\alpha_i\alpha_j=\epsilon_{ijk}\alpha_k^*$, $\alpha_i^*\alpha_j^*=\epsilon_{ijk}\alpha_k$. These additional multiplication rules are not satisfied by the split basis of \cite{furey2016standard}.}
\begin{eqnarray}\label{mtis}
\left\lbrace \alpha_i,\alpha_j \right\rbrace =0,\qquad \left\lbrace \alpha_i^{\dagger},\alpha_j^{\dagger} \right\rbrace =0,\qquad \left\lbrace \alpha_i,\alpha_j^{\dagger} \right\rbrace =\delta_{ij}.
\end{eqnarray}
If one now considers these basis vector operators as elements of $(\mathbb{C}\otimes\mathbb{O})_L\cong C\ell(6)$, then one can construct the primitive idempotent $\omega\omega^{\dagger}=\alpha_1\alpha_2\alpha_3\alpha_3^{\dagger}\alpha_2^{\dagger}\alpha_1^{\dagger}$. The first minimal left ideal is then given by $S^u\equiv C\ell(6)\omega\omega^{\dagger}$. Explicitly:
\begin{eqnarray}
\nonumber S^u\equiv &{}&\\
\nonumber &{}&\;\;\nu \omega\omega^{\dagger}+\\
\nonumber \bar{d}^r\alpha_1^{\dagger}\omega\omega^{\dagger} &+& \bar{d}^g\alpha_2^{\dagger}\omega\omega^{\dagger} + \bar{d}^b\alpha_3^{\dagger}\omega\omega^{\dagger}\\
\nonumber u^r\alpha_3^{\dagger}\alpha_2^{\dagger}\omega\omega^{\dagger} &+& u^g\alpha_1^{\dagger}\alpha_3^{\dagger}\omega\omega^{\dagger} + u^b\alpha_2^{\dagger}\alpha_1^{\dagger}\omega\omega^{\dagger}\\
&+& e^{+}\alpha_3^{\dagger}\alpha_2^{\dagger}\alpha_1^{\dagger}\omega\omega^{\dagger},
\end{eqnarray}
where $\nu$, $\bar{d}^r$ etc. are suggestively labeled complex coefficients denoting the isospin-up elementary fermions. The conjugate system analogously gives a second linearly independent minimal left ideal of isospin-down elementary fermions:
\begin{eqnarray}
\nonumber S^d\equiv &{}&\\
\nonumber &{}&\;\;\bar{\nu} \omega^{\dagger}\omega+\\
\nonumber d^r\alpha_1\omega^{\dagger}\omega &+& d^g\alpha_2\omega^{\dagger}\omega + d^b\alpha_3\omega^{\dagger}\omega\\
\nonumber \bar{u}^r\alpha_3\alpha_2\omega^{\dagger}\omega &+& \bar{u}^g\alpha_1\alpha_3\omega^{\dagger}\omega + \bar{u}^b\alpha_2\alpha_1\omega^{\dagger}\omega\\
&+& e^{-}\alpha_3\alpha_2\alpha_1\omega^{\dagger}\omega.
\end{eqnarray}

One can show that these representations of the minimal left ideals are invariant under the color and electromagnetic symmetries $SU(3)_c$ and $U(1)_{em}$, and each of the basis states in the ideals transforms as a specific lepton or quark under these symmetries as indicated by their suggestively labeled complex coefficients. 

In terms of the Witt basis ladder operators, the $SU(3)$ generators take the form
\begin{eqnarray}
\nonumber\Lambda_1&=&-\alpha_2^{\dagger}\alpha_1-\alpha_1^{\dagger}\alpha_2,\qquad \Lambda_2=i\alpha_2^{\dagger}\alpha_1-i\alpha_1^{\dagger}\alpha_2,\\
\nonumber\Lambda_3&=&\alpha_2^{\dagger}\alpha_2-\alpha_1^{\dagger}\alpha_1,\qquad \Lambda_4=-\alpha_1^{\dagger}\alpha_3-\alpha_3^{\dagger}\alpha_1,\\
\Lambda_5&=&-i\alpha_1^{\dagger}\alpha_3+i\alpha_3^{\dagger}\alpha_1,\qquad \Lambda_6=-\alpha_3^{\dagger}\alpha_2-\alpha_2^{\dagger}\alpha_3,\\
\nonumber\Lambda_7&=&i\alpha_3^{\dagger}\alpha_2-i\alpha_2^{\dagger}\alpha_3,\qquad \Lambda_8=\frac{-1}{\sqrt{3}}(\alpha_1^{\dagger}\alpha_1+\alpha_2^{\dagger}\alpha_2-2\alpha_3^{\dagger}\alpha_3).
\end{eqnarray}
The $U(1)$ generator can be expressed in terms of the Witt basis ladder operators as
\begin{eqnarray}
Q=\frac{1}{3}(\alpha_1^{\dagger}\alpha_1+\alpha_2^{\dagger}\alpha_2+\alpha_3^{\dagger}\alpha_3).
\end{eqnarray}

As an illustrative example we consider $[\Lambda_1, u^g]$:
\begin{eqnarray}
\nonumber\left[ \Lambda_1, u^g\right] &=&\left(-\alpha_2^{\dagger}\alpha_1-\alpha_1^{\dagger}\alpha_2\right) \alpha_1^{\dagger}\alpha_3^{\dagger}-\alpha_1^{\dagger}\alpha_3^{\dagger}\left(-\alpha_2^{\dagger}\alpha_1-\alpha_1^{\dagger}\alpha_2\right),\\
\nonumber &=& -\alpha_2^{\dagger}\alpha_3^{\dagger}\alpha_1\alpha_1^{\dagger}+\alpha_3^{\dagger}\alpha_2^{\dagger}\alpha_1^{\dagger}\alpha_1,\\
\nonumber &=&\alpha_3^{\dagger}\alpha_2^{\dagger}\left( \alpha_1\alpha_1^{\dagger}+\alpha_1^{\dagger}\alpha_1\right) ,\\
&=&\alpha_3^{\dagger}\alpha_2^{\dagger}=u^r.
\end{eqnarray}

%%%%%%%%%%%%%%%%%%%%%%%%%%%%%%%%%%%%%%%%%%%%%%%%%%%%%%%%%%%%%%%%
\section{Three generations of leptons from three  minimal left ideals of $(\mathbb{C}\otimes\mathbb{H})_L\cong C\ell(2)\subset C\ell(6)$}\label{Hideals}

Instead of defining the MTIS of equation (\ref{mtis}), one can instead define the six totally isotropic (but not maximal) subspaces 
\begin{eqnarray}
\left\lbrace \alpha_i^{\dagger}\right\rbrace ,\; \left\lbrace \alpha_i\right\rbrace ,\quad i=1,2,3.
\end{eqnarray}
Each of these totally isotropic subspaces is a MTIS not of the full $C\ell(6)$, but rather of a $C\ell(2)$ subalgebra of $C\ell(6)$. Note that $C\ell(2)\cong(\mathbb{C}\otimes\mathbb{H})_L$, and so each $C\ell(2)$ subalgebra is generated from a $\mathbb{C}\otimes\mathbb{H}$ subalgebra of $\mathbb{C}\otimes\mathbb{O}$. Associated with these subspaces one can define six primitive idempotents
\begin{eqnarray}
\omega_1\omega_1^{\dagger}=\alpha_1\alpha_1^{\dagger},\qquad \omega_2\omega_2^{\dagger}=\alpha_2\alpha_2^{\dagger},\qquad \omega_3\omega_3^{\dagger}=\alpha_3\alpha_3^{\dagger},
\end{eqnarray}
together with their conjugates, and subsequently three ideals
\begin{eqnarray}
\nu_{e}\omega_1\omega_1^{\dagger}+ e^+\alpha_1^{\dagger}\omega_1\omega_1^{\dagger},\quad \nu_{\mu}\omega_2\omega_2^{\dagger}+ \mu^+\alpha_2^{\dagger}\omega_2\omega_2^{\dagger},\quad \nu_{\tau}\omega_3\omega_3^{\dagger}+ \tau^+\alpha_3^{\dagger}\omega_3\omega_3^{\dagger},
\end{eqnarray}
with their conjugates. Here as before, the suggestively labeled complex coefficients denote the isospin-up leptons. The twelve states of these six ideals transform as three generations of leptons with $U(1)_{em}$ symmetry. The $U(1)_{em}$ generator is given by\footnote{In the present case we can define a single $U(1)$ generator $Q$ which acts on all three generations. This will not work in general, where one should instead define one $U(1)_{em}$ generator for each generation.}
\begin{eqnarray}
Q=\alpha_1\alpha_1^{\dagger}+\alpha_2\alpha_2^{\dagger}+\alpha_3\alpha_3^{\dagger}.
\end{eqnarray}

Starting from $(\mathbb{C}\otimes\mathbb{O})_L\cong C\ell(6)$, one can represent either a single generation of leptons and quarks with $SU(3)_c\times U(1)_{em}$ symmetry, or alternatively three generations of leptons with $U(1)_{em}$ symmetry. The former results from constructing the minimal left ideals of $C\ell(6)$, whereas the latter corresponds to constructing the minimal left ideals of three (in this case independent) $C\ell(2)$ subalgebras of $C\ell(6)$. Choosing a preferred octonionic unit $e_7$ singles out three natural quaternionic subalgebras of $\mathbb{O}$ which contain $e_7$. These may be identified as describing three generations of leptons.

This same description of three generations of leptons was considered in \cite{manogue1999dimensional,dray2000quaternionic}, although there these three generations emerged from a dimensional reduction of an octonionic massless Dirac equation in ten dimensions to four dimensions. This reduction likewise leads to three quaternionic subalgebras, singled out because they contain $e_7$, each identified as describing a generation of (massive and massless) leptons. In the next section we demonstrate that the same approach for the sedenions singles out three octonionic algebras, each of which describes a generation of leptons and quarks with electrocolour symmetry.

%%%%%%%%%%%%%%%%%%%%%%%%%%%%%%%%%%%%%%%%%%%%%%%%%%%%%%%%%%%%%%%%%%%%%%%%%%%%%
\section{From octonions to sedenions: one generation to three generations}\label{3generations}

The previous section reviewed earlier results that a single generation of fermions with unbroken $SU(3)_c\times U(1)_{em}$ gauge symmetry can be described starting from $\mathbb{C}\otimes\mathbb{O}$. Whereas $\mathbb{C}$ and $\mathbb{O}$ are each individually division algebras, the tensor product $\mathbb{C}\otimes\mathbb{O}$ is not. Nor is $C\ell(6)$, the adjoint algebra of left actions of $\mathbb{C}\otimes\mathbb{O}$ on itself. Indeed, the construction of minimal left ideals is very dependent on defining a nilpotent basis of ladder operators. The zero divisors in $\mathbb{C}\otimes\mathbb{O}$ therefore play an important role.

The division algebras $\mathbb{C},\mathbb{H},\mathbb{O}$ can each be generated via the Cayley-Dickson process. This process however does not terminate with the octonions, but can be used to generate a $2^n$-dimensional algebra for any positive integer $n$. For $n>3$ however, the Cayley-Dickson algebras are no longer division algebras. However, given the observations above, this should not dissuade us from considering larger Cayley-Dickson algebras. 

%%%%%%%%%%%%%%%%%%%%%%%%%%%%%%%%%%%%%%%%%%%%%%%%
\subsection{The sedenion algebra}

The 16-dimensional sedenion algebra is the fifth Cayley-Dickson algebra $\mathbb{A}_4=\mathbb{S}$. The algebra is not a division algebra, and is non-commutative, non-associative, and non-alternative. However it is power-associative and distributive.

One can choose a canonical basis for $\mathbb{S}$ to be $E_{16}=\lbrace e_i\in\mathbb{S}\vert i=0,1,...,15\rbrace$ where $e_0$ is the real unit and $e_1,...,e_{15}$ are mutually anticommuting imaginary units. In this canonical basis, a general element $a\in\mathbb{S}$ is written as
\begin{eqnarray}
A= \sum_{i=0}^{15} a_i  e_i= a_0+\sum_{i=1}^{15} a_i  e_i,\qquad a_i\in\mathbb{R}.
\end{eqnarray}
The basis elements satisfy the following multiplication rules
\begin{eqnarray}
\nonumber e_0&=&1,\qquad e_0e_i=e_ie_0=e_i,\\
e_1^2&=&e_2^2=...=e_{15}^2=-1,\\
\nonumber e_ie_j&=&-\delta_{ij}e_0+\gamma^k_{ij}e_k,
\end{eqnarray}
where the real structure constants $\gamma^k_{ij}$ are completely antisymmetric. The multiplication table of the sedenion base elements is given in Appendix A. For two sedenions $A,B$, one has
\begin{eqnarray}
AB=\left( \sum_{i=0}^{15} a_i  e_i \right) \left( \sum_{i=0}^{15} b_j  e_i\right) =\sum_{i,j=0}^{15} a_ib_j  (e_ie_j)=\sum_{i,j,k=0}^{15} f_{ij}\gamma^k_{ij}  e_k,
\end{eqnarray}
where $f_{ij}\equiv a_ib_j$. Addition and subtraction is componentwise.

Because the sedenion algebra is not a division algebra, it contains zero divisors. For $\mathbb{S}$ these are elements of the form
\begin{eqnarray}
(e_a+e_b)\circ(e_c+e_b)=0,\qquad e_a,e_b,e_c,e_d\in\mathbb{S}.
\end{eqnarray}
There are 84 such zero divisors, and the subspace of zero divisors of unit norm is homeomorphic to $G_2$ \cite{moreno1997zero}.

%%%%%%%%%%%%%%%%%%%%%%%%%%%%%%%%%%%%%%%%%%%%%%%%%%%%%%%%%%%%%%%%%%%%%%%%%%%%%%%%%%%%%%%
%\subsection{Endomorphisms and left adjoint algebras of Cayley-Dickson algebras}

%%%%%%%%%%%%%%%%%%%%%%%%%%%%%%%%%%%%%%%%%%%%%%%%%%%%%%%%%%%
\subsection{Why sedenions?}

One appealing reason for considering division algebras in relation to particle symmetries (both spacetime and internal) is that, unlike Lie algebras and Clifford algebras, there are only a finite number of division algebras\footnote{There are many convincing arguments that the four division algebras should play a special role in particle physics. Many of these arguments can be found in the extensive works of Dixon \cite{dixon2013division,dixon2017division}.}. At the same time, each division algebra generates very specific Lie and Clifford algebras. Starting with a division algebra, the physical symmetries are dictated. 

Unlike the four division algebras, the sedenions have not been the focus of much study in relation to particle physics, presumably due to the fact that they are not a division algebra and are neither associative nor alternative. Tensor products of division algebras however also contain zero divisors. The constructions of particle symmetries from $\mathbb{C}\otimes\mathbb{O}$ \cite{furey2016standard}, $C\ell(6)$ \cite{Stoica2018}, and $\mathbb{T}=\mathbb{R}\otimes\mathbb{C}\otimes\mathbb{H}\otimes\mathbb{O}$ \cite{dixon2013division}, all contain zero divisors, and these play an important role. It seems reasonable therefore to continue the Cayley-Dickson algebra into the non-division algebras.

The symmetries of these algebras can be extracted from their automorphism groups. Interestingly, for the sedenions one finds that
\begin{eqnarray}\label{autS}
\mathrm{Aut}(\mathbb{S})=\mathrm{Aut}(\mathbb{O})\times S_3.
\end{eqnarray}
The only difference between the octonions and sedenion automorphism groups is a factor of the permutation group $S_3$. This permutation group can be constructed from the triality automorphism of $\mathrm{Spin}(8)$, the spin group of $C\ell(8)$ \cite{lounesto2001caa}. This automorphism is of order three. Interestingly then, what Eq.(\ref{autS}) suggests is that the fundamental symmetries of $\mathbb{S}$ are the same as those of $\mathbb{O}$, although the factor of $S_3$ suggests we get three copies. This is exactly what we want in order to describe the observed three generations of fermions.

For higher Cayley-Dickson ($n>3$) algebras one has
\begin{eqnarray}
\mathrm{Aut}(\mathbb{A}_n)=\mathrm{Aut}(\mathbb{O})\times (n-3)S_3.
\end{eqnarray}
This tells us that the underlying symmetries are $G_2$, the automorphism group of the octonions. The higher Cayley-Dickson algebras only add additional trialities, and perhaps no new fundamental physics should be expected beyond $\mathbb{C}\otimes\mathbb{S}$.

%The set of all linear transformations of the complex sedenions generated the algebra of $16\times 16$ complex matrices, $\mathbb{C}(16)$. This matrix algebra faithfully represents the Clifford algebra $C\ell(8)$, and so we have that $\mathrm{End}(\mathbb{C}\otimes\mathbb{S})\cong \mathbb{C}(16)\cong C\ell(8)$. Equivalently, the adjoint algebras of larger Cayley-Dickson algebras will be larger than $C\ell(8)$, with $\mathrm{End}(\mathbb{C}\otimes\mathbb{A}_n)\cong C\ell(2n)$. However Bott periodicity means that $C\ell(0,n+8)=C\ell(0,8)\times C\ell(0,n)$, so for larger algebras, one expects just additional copies of $C\ell(8)$.

%%%%%%%%%%%%%%%%%%%%%%%%%%%%%%%%%%%%%%%%%%%%%%%%%%%%%%%%%%%%%%%%%%%%%%%%%%%%%
\subsection{From one generation to three generations}

A natural way to extend the intriguing results from Section (\ref{octonions}) beyond a single generation of elementary fermions is to consider the complex sedenions $\mathbb{C}\otimes\mathbb{S}$ instead of complex octonions $\mathbb{C}\otimes\mathbb{O}$. This larger algebra also generates a larger adjoint algebra of left actions $(\mathbb{C}\otimes\mathbb{S})_L$. The key point of this paper is to demonstrate:
\begin{eqnarray}
\nonumber \mathbb{C}\otimes\mathbb{O} &\rightarrow & \mathbb{C}\otimes\mathbb{S}\\
\nonumber (\mathbb{C}\otimes\mathbb{O})_L&\rightarrow &(\mathbb{C}\otimes\mathbb{S})_L\\
\nonumber \textrm{1\;generation}&\rightarrow & \textrm{3\; generations}.
\end{eqnarray}

\begin{figure}
\centering
\includegraphics[width=0.33\linewidth]{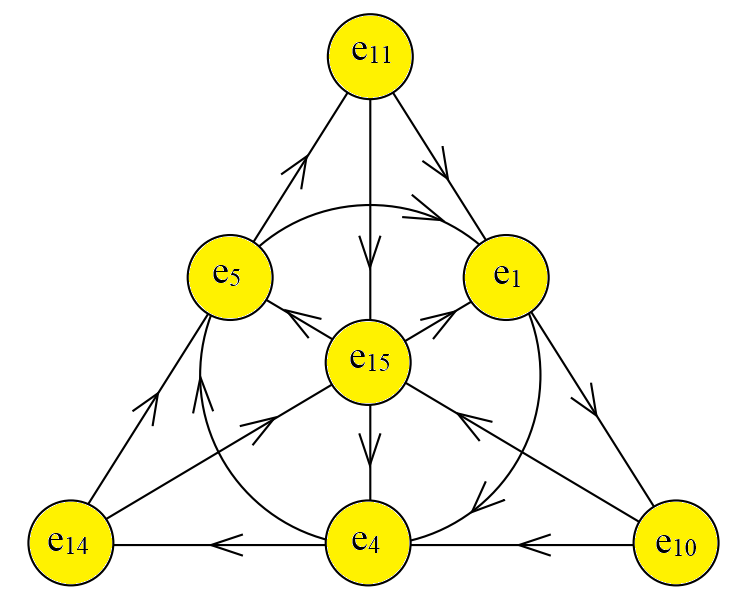}\label{Fano1}
\qquad
\includegraphics[width=0.33\linewidth]{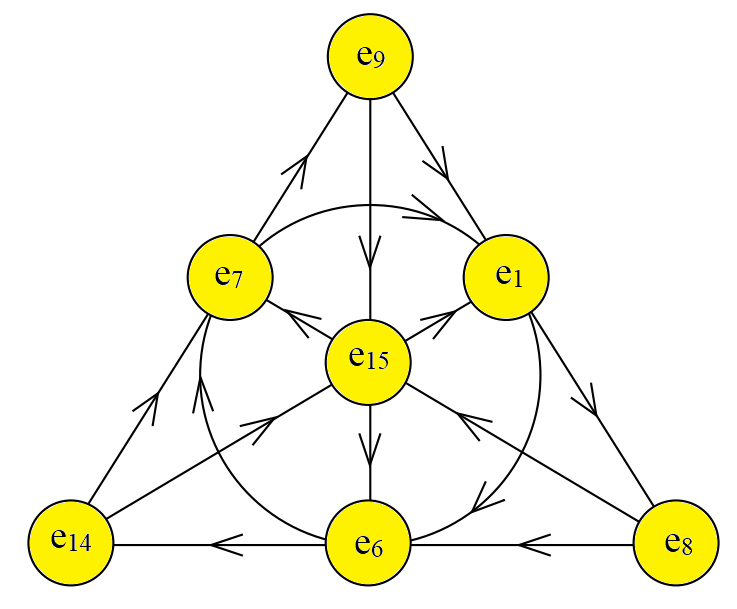}\label{Fano2}
\qquad
\includegraphics[width=0.33\linewidth]{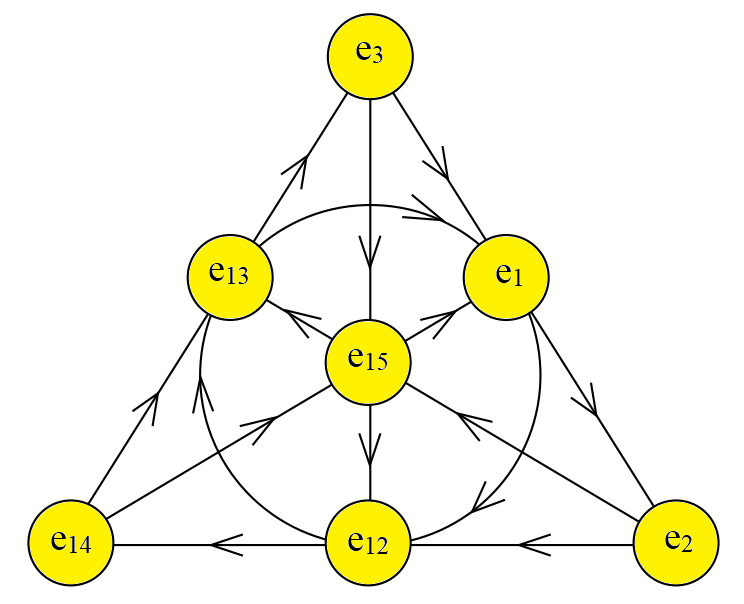}\label{Fano3}
\caption{Fano planes of three octonionic subalgebras in the sedenions.}
\label{Fano}
\end{figure}

As in section (\ref{octonions}), we start by choosing $e_{15}$ as a special imaginary unit from the sedenion basis, mimicking the common choice of $e_7$ for $\mathbb{O}$. One can then define the primitive idempotents
\begin{eqnarray}
\rho_+=\frac{1}{2}(1+ie_{15}),\qquad \rho_-=\frac{1}{2}(1-ie_{15}),
\end{eqnarray}
satisfying
\begin{eqnarray}
\rho_++\rho_-=1,\qquad \rho_{\pm}^2=\rho_{\pm},\qquad \rho_+\rho_-=0.
\end{eqnarray}  
We now act with $\rho_+$ on the sedenion basis elements $\lbrace e_1,...,e_{14}\rbrace$, and subsequently define a split basis for $\mathbb{C}\otimes\mathbb{S}$. This gives a set of seven nilpotents $\eta_i$, $i=1,2,...,7$ of the form $\frac{1}{2}(-e_a+ ie_{15-a})$ where $a\in{5,7,13,1,4,6,12}$, together with seven nilpotent conjugates $\eta_i^{\dagger}$ of the form $\frac{1}{2}(e_a+ ie_{15-a})$. Considered as elements of $(\mathbb{C}\otimes\mathbb{S})_L$, these define two MTIS satisfying
\begin{eqnarray}
\left\lbrace \eta_i^{\dagger},\eta_j^{\dagger}\right\rbrace =\left\lbrace \eta_i,\eta_j\right\rbrace =0,\qquad \left\lbrace \eta_i^{\dagger},\eta_j\right\rbrace =\delta_{ij}.
\end{eqnarray}
These $\eta_i^{\dagger}$ and $\eta_i$ uniquely divide into three sets of ladder operators that do 
\begin{eqnarray}
\alpha_1^{\dagger}&=&\frac{1}{2}(e_5+ie_{10}),\qquad \beta_1^{\dagger}=\frac{1}{2}(e_7+ie_8),\qquad \gamma_1^{\dagger}=\frac{1}{2}(e_{13}+ie_2),\\
\alpha_2^{\dagger}&=&\beta_2^{\dagger}=\gamma_2^{\dagger}=\frac{1}{2}(e_1+ie_{14}),\\
\alpha_3^{\dagger}&=&\frac{1}{2}(e_4+ie_{11}),\qquad \beta_3^{\dagger}=\frac{1}{2}(e_6+ie_9),\qquad \gamma_3^{\dagger}=\frac{1}{2}(e_{12}+ie_3),
\end{eqnarray}
together with
\begin{eqnarray}
\alpha_1&=&\frac{1}{2}(-e_5+ie_{10}),\qquad \beta_1=\frac{1}{2}(-e_7+ie_8),\qquad \gamma_1=\frac{1}{2}(-e_{13}+ie_2),\\
\alpha_2&=&\beta_2=\gamma_2=\frac{1}{2}(-e_1+ie_{14}),\\
\alpha_3&=&\frac{1}{2}(-e_4+ie_{11}),\qquad \beta_3=\frac{1}{2}(-e_6+ie_9),\qquad \gamma_3=\frac{1}{2}(-e_{12}+ie_3).
\end{eqnarray}
Here, $\lbrace \alpha_i^{\dagger},\alpha_i\rbrace$,  $\lbrace \beta_i^{\dagger},\beta_i\rbrace$, and $\lbrace \gamma_i^{\dagger},\gamma_i\rbrace$ each individually form a split basis for $\mathbb{C}\otimes\mathbb{O}$ and satisfy the anticommutator algebra of fermionic ladder operators exactly as in section (\ref{octonions})\footnote{There is a slight deviation here in the definition of ladder operators as in \cite{furey2016standard}. Here we have multiplied the standard basis elements $e_i$ on the left by $\rho_+$ and from the result have immediately identified $\eta_i$ and $\eta_i^{\dagger}$. This means that the real parts $\alpha_i^{\dagger}$ (and similarly for $\beta_i^{\dagger}$ and $\gamma_i^{\dagger}$) form a quaternion subalgebra. In \cite{furey2016standard} it is the imaginary parts that form a quaternion subalgebra.}. By uniquely we mean that any other division of the seven $\eta_i^{\dagger}$ and seven $\eta_i$ into equally sized sets does not result in three $\mathbb{C}\otimes\mathbb{O}$ subalgebras. Thus, what we find is that the action of the projector $\rho_+$ on the standard sedenion basis uniquely divides the algebra into three sets of split basis elements for $\mathbb{C}\otimes\mathbb{O}$. This reflects what happened earlier when considering $\mathbb{C}\otimes\mathbb{O}$. There, choosing $e_7$ as the special unit imaginary singles out three quaternionic subalgebras. In \cite{manogue1999dimensional,dray2000quaternionic} these are identified with three generations of leptons. Instead here we have found that selecting $e_{15}$ as the special unit imaginary singles our three octonionic subalgebras. This allows us to describe three generations of leptons and quarks with unbroken $SU(3)_c\times U(1)_{em}$ symmetry. These three $\mathbb{C}\otimes\mathbb{O}$ subalgebras are however not independent. All three subalgebras share a common quaternionic subalgebra spanned by $\lbrace 1, e_1, e_{14}, e_{15}\rbrace$. The three intersecting octonion subalgebras together with the multiplication rules of their base elements may be represented by the three Fano planes in Figure \ref{Fano}.

In addition to the standard fermionic ladder operator algebras satisfied by the $\alpha$'s, $\beta$'s, and $\gamma$'s separately, we also have
\begin{eqnarray}
\nonumber \lbrace \alpha_i^{\dagger},\beta_j^\dagger\rbrace&=&\lbrace \beta_i^{\dagger},\gamma_j^\dagger\rbrace=\lbrace \gamma_i^{\dagger},\alpha_j^\dagger\rbrace=0,\\
\lbrace \alpha_i,\beta_j\rbrace&=&\lbrace \beta_i,\gamma_j\rbrace=\lbrace \gamma_i,\alpha_j\rbrace=0,\\
\nonumber\lbrace \alpha_i^{\dagger},\beta_j\rbrace&=&\lbrace \beta_i^{\dagger},\gamma_j\rbrace=\lbrace \gamma_i^{\dagger},\alpha_j\rbrace=\delta_{ij}.
\end{eqnarray}

From each of the three $\mathbb{C}\otimes\mathbb{O}$ subalgebras we can now construct an ideal following the procedure or Section \ref{octonions}. These ideals will be minimal left $C\ell(6)$ ideals embedded within the full $(\mathbb{C}\otimes\mathbb{S})_L$ algebra. Using the first set of ladder operators $\lbrace \alpha_1^{\dagger},\alpha_2^{\dagger},\alpha_3^{\dagger},\alpha_1,\alpha_2,\alpha_3\rbrace$, we can now construct the first ideal as $S_1^u=C\ell(6)\omega_1\omega_1^{\dagger}$, where $\omega_1\omega_1^{\dagger}=\alpha_1\alpha_2\alpha_3\alpha_3^{\dagger}\alpha_2^{\dagger}\alpha_1^{\dagger}$ is a $C\ell(6)$ primitive idempotent.
\begin{eqnarray}
\nonumber S_1^u\equiv &{}&\\
\nonumber &{}&\;\;\nu_e \omega_1\omega_1^{\dagger}+\\
\nonumber \bar{d}^r\alpha_1^{\dagger}\omega_1\omega_1^{\dagger} &+& \bar{d}^g\alpha_2^{\dagger}\omega_1\omega_1^{\dagger} + \bar{d}^b\alpha_3^{\dagger}\omega_1\omega_1^{\dagger}\\
\nonumber u^r\alpha_3^{\dagger}\alpha_2^{\dagger}\omega_1\omega_1^{\dagger} &+& u^g\alpha_1^{\dagger}\alpha_3^{\dagger}\omega_1\omega_1^{\dagger} + u^b\alpha_2^{\dagger}\alpha_1^{\dagger}\omega_1\omega_1^{\dagger}\\
&+& e^{+}\alpha_3^{\dagger}\alpha_2^{\dagger}\alpha_1^{\dagger}\omega_1\omega_1^{\dagger},
\end{eqnarray}
The complex conjugate system gives a second linearly independent $C\ell(6)$ minimal left ideal
\begin{eqnarray}
\nonumber S_1^d\equiv &{}&\\
\nonumber &{}&\;\;\bar{\nu}_e \omega_1^{\dagger}\omega_1+\\
\nonumber d^r\alpha_1\omega_1^{\dagger}\omega_1 &+& d^g\alpha_2\omega_1^{\dagger}\omega_1 + d^b\alpha_3\omega_1^{\dagger}\omega_1\\
\nonumber \bar{u}^r\alpha_3\alpha_2\omega_1^{\dagger}\omega_1 &+& \bar{u}^g\alpha_1\alpha_3\omega_1^{\dagger}\omega_1 + \bar{u}^b\alpha_2\alpha_1\omega_1^{\dagger}\omega_1\\
&+& e^{-}\alpha_3\alpha_2\alpha_1\omega_1^{\dagger}\omega_1.
\end{eqnarray}

One can likewise construct the ideals of the other two $C\ell(6)$ subalgebras from the sets of ladder operators $\lbrace \beta_i^{\dagger},\beta_i\rbrace$ and $\lbrace \gamma_i^{\dagger}, \gamma_i\rbrace$. In total we find six minimal left ideals $\lbrace S_1^u, S_1^d, S_2^u, S_2^d, S_3^u, S_3^d\rbrace$. Whereas $S_i^u$ and $S_i^d$ are linearly independent, $S_i^u$ and $S_j^u$ are not.

%%%%%%%%%%%%%%%%%%%%%%%%%%%%%%%%%%%%%%%%%%%%%%%%%%%%%%%%%%%%%%%%%%%%%
\section{Unbroken $SU(3)_c\times U(1)_{em}$ gauge symmetries}\label{symmetries}

As for the case of a single generation constructed from $\mathbb{C}\otimes\mathbb{O}$, the physical states of the three generation extension presented here transform as three generations of SM fermions under $SU(3)_c\times U(1)_{em}$. Each generation has its own copy of $SU(3)$ and $U(1)$, which we label $SU^{(i)}(3)$ and $U^{(i)}(1)$, $i=1,2,3$.

The $SU^{(1)}(3)$ generators can be represented in the first $C\ell(6)\subset (\mathbb{C}\otimes\mathbb{S})_L$ as
\begin{eqnarray}
\nonumber \Lambda_1^{(1)}&=&-\frac{i}{2}(e_1e_{10}-e_{14}e_5),\qquad \Lambda_2^{(1)}=\frac{i}{2}(e_{14}e_{10}-e_5e_1),\\
\nonumber \Lambda_3^{(1)}&=&-\frac{i}{2}(e_5e_{10}-e_1e_{14}),\qquad \Lambda_4^{(1)}=-\frac{i}{2}(e_5e_{11}-e_{10}e_4),\\
\Lambda_5^{(1)}&=&-\frac{i}{2}(e_{10}e_{11}-e_4e_5),\qquad \Lambda_6^{(1)}=-\frac{i}{2}(e_4e_{14}-e_{11}e_1),\\
\nonumber\Lambda_7^{(1)} &=& \frac{i}{2}(e_{11}e_{14}-e_1e_4),\qquad\;\;\; \Lambda_8^{(1)}=\frac{-i}{\sqrt{3}}(e_5e_{10}+e_1e_{14}-2e_4e_{11}).
\end{eqnarray}
This can be rewritten in terms of the ladder operators, taking the form
\begin{eqnarray}
\nonumber\Lambda_1^{(1)}&=&-(\alpha_2^{\dagger}\alpha_1+\alpha_1^{\dagger}\alpha_2),\qquad\quad\;\; \Lambda_2^{(1)}=-i(\alpha_2^{\dagger}\alpha_1-\alpha_1^{\dagger}\alpha_2),\\
\nonumber\Lambda_3^{(1)}&=&\alpha_2^{\dagger}\alpha_2-\alpha_1^{\dagger}\alpha_1,\qquad\quad \Lambda_4^{(1)}=-(\alpha_1^{\dagger}\alpha_3+\alpha_3^{\dagger}\alpha_1),\\
\Lambda_5^{(1)}&=&i(\alpha_1^{\dagger}\alpha_3-\alpha_3^{\dagger}\alpha_1),\qquad \Lambda_6^{(1)}=-(\alpha_3^{\dagger}\alpha_2+\alpha_2^{\dagger}\alpha_3),\\
\nonumber\Lambda_7^{(1)}&=&-i(\alpha_3^{\dagger}\alpha_2-\alpha_2^{\dagger}\alpha_3),\qquad\;\;\;\; \Lambda_8^{(1)}=\frac{1}{\sqrt{3}}(-\alpha_1^{\dagger}\alpha_1-\alpha_2^{\dagger}\alpha_2+2\alpha_3^{\dagger}\alpha_3).
\end{eqnarray}
The explicit $SU(3)$ generators for the other two generations are given in Appendix B. Under the action of $SU^{(1)}(3)$, one finds that the ideals $S_1^{u}$ and $S_1^{d}$ transform as \cite{gunaydin1973quark,dixon1990derivation,furey2016standard}
\begin{eqnarray}
S_1^u \sim \mathbf{1}\oplus \bar{\mathbf{3}}\oplus \mathbf{3}\oplus \bar{\mathbf{1}},\qquad S_1^d\sim \bar{\mathbf{1}}\oplus \mathbf{3}\oplus \bar{\mathbf{3}}\oplus \mathbf{1}.
\end{eqnarray}

As two illustrative examples, consider $\left[  \Lambda_1^{(2)}, s^r\right]$ and $\left[ Q^{(3)}, \bar{t}^b\right]$:
\begin{eqnarray}
\nonumber\left[ \Lambda_1^{(2)},s^r\right] &=&-(\beta_2^{\dagger}\beta_1+\beta_1^{\dagger}\beta_2)\beta_1+\beta_1(\beta_2^{\dagger}\beta_1+\beta_1^{\dagger}\beta_2),\\
\nonumber&=& \beta_1^{\dagger}\beta_1\beta_2+\beta_1\beta_1^{\dagger}\beta_2,\\
&=& (\beta_1^{\dagger}\beta_1+\beta_1\beta_1^{\dagger})\beta_2=\beta_2=s^g,\\
\nonumber\left[ Q^{(3)}, \bar{t}^b\right] &=& \frac{1}{3}(\gamma_1^{\dagger}\gamma_1+\gamma_2^{\dagger}\gamma_2+\gamma_3^{\dagger}\gamma_3)\gamma_2\gamma_1-\frac{1}{3}\gamma_2\gamma_1(\gamma_1^{\dagger}\gamma_1+\gamma_2^{\dagger}\gamma_2+\gamma_3^{\dagger}\gamma_3),\\
&=& -\frac{1}{3}\gamma_2\gamma_1\gamma_1^{\dagger}\gamma_1+\frac{1}{3}\nonumber\gamma_1\gamma_2\gamma_2^{\dagger}\gamma_2,\\
\nonumber&=& -\frac{1}{3}\gamma_2\gamma_1(-\gamma_1\gamma_1^{\dagger}+1)+\frac{1}{3}\gamma_1(-\gamma_2^{\dagger}\gamma_2+1)\gamma_2,\\
&=&-\frac{2}{3}\gamma_2\gamma_1=-\frac{2}{3}\bar{t}^b.
\end{eqnarray}

%%%%%%%%%%%%%%%%%%%%%%%%%%%%%%%%%%%%%%%%%%%%%%%%%%%%%%%%%%%%%%%%%%%
\subsection{Mixing between generations}

The ladder operators for each generation in our three generation model belong to a Fano plane. Each Fano plane gives the projective geometry of the octonionic projective plane $OP^2$ \cite{baez2002octonions}, which is the quantum state space upon which the quantum exceptional Jordan algebra $J_3(\mathbb{O})$ acts \cite{manogue2010octonions}. One might therefore expect a relationship between the approach presented here, based on $\mathbb{C}\otimes\mathbb{O})_L$ and three generation models based on $J_3(\mathbb{O})$ \cite{dubois2016exceptional,todorov2018deducing,dubois2019exceptional}.

For the case of three non-intersecting Fano planes, the total state space is given by the tensor product of the three individual state spaces. This would mean that what we called $\Lambda^{(1)}_i$ is really $\Lambda\otimes1\otimes1$, and what we called a blue top quark $t^b$ is really $1\otimes1\otimes t^b$. It follows that any inter-generational commutator such as $[\Lambda_i^{(1)},t^b]$ vanishes trivially. In our construction however, the three Fano planes are not truly independent of one another, but actually share a common complex quaternionic line of intersection. This means that the tensor product structure is not perfectly commutative, and this can lead to mixing between generations. For example
\begin{eqnarray}
\left[ \Lambda_1^{(2)},\alpha_2^{\dagger}\right] =\beta_1^{\dagger}.
\end{eqnarray}
What this curious behaviour corresponds to physically is currently being investigated. One interesting proposal is that the mixing between generations might form a basis for the weak PMNS and CKM mixing matrices\footnote{It may also be that the tensor product gets braided. The relation between braids and the minimal left ideals of $C\ell(6)$ were reported on in \cite{gresnigt2018braids}. Embedding braided matter into Loop Quantum Gravity is possible when the underlying group representations that label the edges of a spin-network are $q$-deformed. Such a $q$-deformation has an associated $R$-matrix that braids any tensor products. This idea is still speculative.}.

%%%%%%%%%%%%%%%%%%%%%%%%%%%%%%%%%%%%%%%%%
\section{Discussion}

The adjoint algebra of left actions of the complex octonions was previously shown to have the right mathematical structure to describe one generation of SM fermions with the two unbroken gauge symmetries $SU(3)_c$ and $U(1)_{em}$ \cite{furey2016standard}. In this paper we have extended these results to exactly three generations by instead starting from the complex sedenions. A split basis for this algebra naturally divides fourteen nilpotent ladder operators into three sets, each of which gives a split basis for a complex octonion subalgebra. Each of these subalgebras generates, via the minimal left ideals of their adjoint algebra of left actions, a single generation of fermions.

The minimal left ideals in this case are minimal left ideals of $(\mathbb{C}\times\mathbb{O})_L\cong C\ell(6)$, but not of the full $(\mathbb{C}\otimes\mathbb{S})_L$. Our construction therefore generalizes the construction outlined in Section \ref{Hideals}. Instead of defining the minimal left ideals, via the MTIS, of the $C\ell(6)$ one can instead divide the split basis for $\mathbb{C}\otimes\mathbb{O}$ into three split bases for three $\mathbb{C}\otimes\mathbb{H}$ subalgebras. These three quaternionic subalgebra all contain the special imaginary unit $e_7$ used to construct the projectors. In earlier works these three quaternionic subalgebras have been identified with three generations of leptons \cite{manogue1999dimensional}. For the case of $\mathbb{C}\otimes\mathbb{S}$, instead of constructing the minimal left ideals of $(\mathbb{C}\otimes\mathbb{S})_L$, we have used the special imaginary unit $e_{15}$ to define three $\mathbb{C}\otimes\mathbb{O}$ subalgebras. These three subalgebras share a common $\mathbb{C}\otimes\mathbb{H}$ algebras.

The sedenions are not a division algebra. Unlike $\mathbb{R,C,H}$ and $\mathbb{O}$ they are not alternative. However the lack of alternativity, like the lack of associativity, does not affect the construction of minimal left ideals. The fermion algebra satisfied by the split basis requires  only two basis elements to be multiplied together at any time. In any such calculation, associativity or alternativity play no role. The ideals are constructed from the adjoint algebra of left actions of the algebra on itself and is always associative, alternative, and isomorphic to a Clifford algebra. 

The sedenions should not be ruled out as playing a role in particle physics on the basis that they do not constitute a division algebra. Whereas $\mathbb{R,C,H}$ and $\mathbb{O}$ are by themselves division algebras, tensor products such as $\mathbb{C}\otimes\mathbb{H}$ and $\mathbb{C}\otimes\mathbb{O}$ are not, and in fact the zero divisors of these algebra play a crucial role in the constructions of left ideals \cite{furey2016standard,dixon2013division,dixon2010division}. Because $\textrm{Aut}(\mathbb{S})=\textrm{Aut}(\mathbb{O})\times S_3$, the fundamental symmetries of $\mathbb{S}$ are the same as those of $\mathbb{O}$. The factor of $S_3$ is generated by the triality automorphism of $\textrm{Spin}(8)$. In a future paper the relation between $\mathbb{S}$, triality, and $\textrm{Spin}(8)$ will be presented in more detail. The combination of the octonion algebra and triality leads naturally to the exceptional Jordan algebra, a 27-dimensional quantum algebra. This should make it possible to relate our work to the exceptional Jordan algebra \cite{dubois2019exceptional,todorov2018deducing}. Beyond the sedenions, one might not expect new physics. This is because $\textrm{Aut}(\mathbb{A}_n)=\textrm{Aut}(\mathbb{O})\times (n-3)S_3$, where $\mathbb{A}_n$ is the $n$-th Cayley-Dickson algebra. The fundamental symmetries remain those of the octonions, with only additional factors of the permutation group $S_3$ being added. 

Finally, The three copies of $(\mathbb{C}\otimes\mathbb{O})_L\cong C\ell(6)$ live inside the larger algebra $(\mathbb{C}\otimes\mathbb{S})_L$. This larger algebra contains additional structure, although it is not clear yet what this corresponds to physically. One interesting idea is that it may form a theoretical basis for the PMNS and CKM matrices that describe neutrino oscillations and quark mixing. This idea remains to be investigated.

%%%%%%%%%%%%%%%%%%%%%%%%%%%%%%%%%%%%%%%%%%%%%%%%%%%%%
\section*{Appendix A: Sedenion multiplication table}

\tiny
\begin{center}
 \begin{tabular}{|c || c| c| c| c| c| c| c| c| c| c| c| c| c| c| c|}
 \hline
 $e_0$ & $e_1$ & $e_2$ & $e_3$ & $e_4$ & $e_5$ & $e_6$ & $e_7$ & $e_8$ & $e_9$ & $e_{10}$ & $e_{11}$ & $e_{12}$ & $e_{13}$ & $e_{14}$ & $e_{15}$ \\ [0.5ex] 
 \hline\hline
 $e_1$ & $-e_0$ &$e_3$ & $-e_2$ & $e_5$ & $-e_4$ & $e_7$ & $-e_6$ & $e_9$ & $-e_8$ & $e_{11}$ & $-e_{10}$ & $e_{13}$ & $-e_{12}$ & $e_{15}$ & $-e_{14}$ \\
 \hline
 $e_2$ & $-e_3$ &$-e_0$ & $e_1$ & $e_6$ & $-e_7$ & $-e_4$ & $e_5$ & $e_{10}$ & $-e_{11}$ & $-e_{8}$ & $e_{9}$ & $e_{14}$ & $-e_{15}$ & $-e_{12}$ & $e_{13}$ \\
 \hline
  $e_3$ & $e_2$ &$-e_1$ & $-e_0$ & $-e_8$ & $-e_6$ & $e_5$ & $e_4$ & $-e_{11}$ & $-e_{10}$ & $e_{9}$ & $e_{8}$ & $-e_{15}$ & $-e_{14}$ & $e_{13}$ & $e_{12}$ \\
  \hline
   $e_4$ & $-e_5$ &$-e_6$ & $e_7$ & $-e_0$ & $e_1$ & $e_2$ & $-e_3$ & $e_{12}$ & $-e_{13}$ & $-e_{14}$ & $e_{15}$ & $-e_{8}$ & $e_{9}$ & $e_{10}$ & $-e_{11}$ \\  
     \hline
   $e_5$ & $e_4$ &$e_7$ & $e_6$ & $-e_1$ & $-e_0$ & $-e_3$ & $-e_2$ & $-e_{13}$ & $-e_{12}$ & $e_{15}$ & $e_{14}$ & $e_{9}$ & $e_{8}$ & $-e_{11}$ & $-e_{10}$ \\  
       \hline
   $e_6$ & $-e_7$ &$e_4$ & $-e_5$ & $-e_2$ & $e_3$ & $-e_0$ & $e_1$ & $-e_{14}$ & $e_{15}$ & $-e_{12}$ & $e_{13}$ & $e_{10}$ & $-e_{11}$ & $e_{8}$ & $-e_{9}$ \\
       \hline
   $e_7$ & $e_6$ &$-e_5$ & $-e_4$ & $e_3$ & $e_2$ & $-e_1$ & $-e_0$ & $e_{15}$ & $e_{14}$ & $e_{13}$ & $e_{12}$ & $-e_{11}$ & $-e_{10}$ & $-e_{9}$ & $-e_{8}$ \\
       \hline
   $e_8$ & $-e_9$ &$-e_{10}$ & $e_{11}$ & $-e_{12}$ & $e_{13}$ & $e_{14}$ & $-e_{15}$ & $-e_{0}$ & $e_{1}$ & $e_{2}$ & $-e_{3}$ & $e_{4}$ & $-e_{5}$ & $-e_{6}$ & $e_{7}$ \\  
          \hline
   $e_9$ & $e_8$ &$e_{11}$ & $e_{10}$ & $e_{13}$ & $e_{12}$ & $-e_{15}$ & $-e_{14}$ & $-e_{1}$ & $-e_{0}$ & $-e_{3}$ & $-e_{2}$ & $-e_{5}$ & $-e_{4}$ & $e_{7}$ & $e_{6}$ \\  
           \hline
   $e_{10}$ & $-e_{11}$ &$e_{8}$ & $-e_{9}$ & $e_{14}$ & $-e_{15}$ & $e_{12}$ & $-e_{13}$ & $-e_{2}$ & $e_{3}$ & $-e_{0}$ & $e_{1}$ & $-e_{6}$ & $e_{7}$ & $-e_{4}$ & $e_{5}$ \\
           \hline
   $e_{11}$ & $e_{10}$ &$-e_{9}$ & $-e_{8}$ & $-e_{15}$ & $-e_{14}$ & $-e_{13}$ & $-e_{12}$ & $e_{3}$ & $e_{2}$ & $-e_{1}$ & $-e_{0}$ & $e_{7}$ & $e_{6}$ & $e_{5}$ & $e_{4}$ \\   
              \hline
   $e_{12}$ & $-e_{13}$ &$-e_{14}$ & $e_{15}$ & $e_{8}$ & $-e_{9}$ & $-e_{10}$ & $e_{11}$ & $-e_{4}$ & $e_{5}$ & $e_{6}$ & $-e_{7}$ & $-e_{0}$ & $e_{1}$ & $e_{2}$ & $-e_{3}$ \\ 
    \hline
   $e_{13}$ & $e_{12}$ &$e_{15}$ & $e_{14}$ & $-e_{9}$ & $-e_{8}$ & $e_{11}$ & $e_{10}$ & $e_{5}$ & $e_{4}$ & $-e_{7}$ & $-e_{6}$ & $-e_{1}$ & $-e_{0}$ & $-e_{3}$ & $-e_{2}$ \\ 
    \hline
   $e_{14}$ & $-e_{15}$ &$e_{12}$ & $-e_{13}$ & $-e_{10}$ & $e_{11}$ & $-e_{8}$ & $e_{9}$ & $e_{6}$ & $-e_{7}$ & $e_{4}$ & $-e_{5}$ & $-e_{2}$ & $e_{3}$ & $-e_{0}$ & $e_{1}$ \\
    \hline
   $e_{15}$ & $e_{14}$ &$-e_{13}$ & $-e_{12}$ & $e_{11}$ & $e_{10}$ & $e_{9}$ & $e_{8}$ & $-e_{7}$ & $-e_{6}$ & $-e_{5}$ & $-e_{4}$ & $e_{3}$ & $e_{2}$ & $-e_{1}$ & $-e_{0}$ \\  
 \hline

\end{tabular}
\end{center}
\normalsize

%%%%%%%%%%%%%%%%%%%%%%%%%%%%%%%%%%%%%%%%%%%%%%%%%%%%%
\section*{Appendix B: $C\ell(6)$ minimal left ideals in $(\mathbb{C}\otimes\mathbb{S})_L$}

%%%%%%%%%%%%%%%%%%%%%%%%%%%%%%%%%%
\subsection*{Generation 1}

\begin{eqnarray}
\nonumber\alpha_1^{\dagger}&=&\frac{1}{2}(e_5+ie_{10}),\qquad \alpha_2^{\dagger}=\frac{1}{2}(e_1+ie_{14}),\qquad \alpha_3^{\dagger}=\frac{1}{2}(e_4+ie_{11}),\\
\alpha_1&=&\frac{1}{2}(-e_5+ie_{10}),\qquad \alpha_2=\frac{1}{2}(-e_1+ie_{14}),\qquad \alpha_3=\frac{1}{2}(-e_4+ie_{11}),\\
\nonumber \omega_1\omega_1^{\dagger}&=&\alpha_1\alpha_2\alpha_3\alpha_3^{\dagger}\alpha_2^{\dagger}\alpha_1^{\dagger}.
\end{eqnarray}

The first minimal left ideal is given by $S_1^u\equiv C\ell(6)_1\omega_1\omega_1^{\dagger}$. Explicitly,
\begin{eqnarray}
\nonumber S_1^u\equiv &{}&\\
\nonumber &{}&\;\;\nu_e \omega_1\omega_1^{\dagger}+\\
\nonumber \bar{d}^r\alpha_1^{\dagger}\omega_1\omega_1^{\dagger} &+& \bar{d}^g\alpha_2^{\dagger}\omega_1\omega_1^{\dagger} + \bar{d}^b\alpha_3^{\dagger}\omega_1\omega_1^{\dagger}\\
\nonumber u^r\alpha_3^{\dagger}\alpha_2^{\dagger}\omega_1\omega_1^{\dagger} &+& u^g\alpha_1^{\dagger}\alpha_3^{\dagger}\omega_1\omega_1^{\dagger} + u^b\alpha_2^{\dagger}\alpha_1^{\dagger}\omega_1\omega_1^{\dagger}\\
&+& e^{+}\alpha_3^{\dagger}\alpha_2^{\dagger}\alpha_1^{\dagger}\omega_1\omega_1^{\dagger},
\end{eqnarray}
The complex conjugate system gives a second linearly independent minimal left ideal
\begin{eqnarray}
\nonumber S_1^d\equiv &{}&\\
\nonumber &{}&\;\;\bar{\nu}_e \omega_1^{\dagger}\omega_1+\\
\nonumber d^r\alpha_1\omega_1^{\dagger}\omega_1 &+& d^g\alpha_2\omega_1^{\dagger}\omega_1 + d^b\alpha_3\omega_1^{\dagger}\omega_1\\
\nonumber \bar{u}^r\alpha_3\alpha_2\omega_1^{\dagger}\omega_1 &+& \bar{u}^g\alpha_1\alpha_3\omega_1^{\dagger}\omega_1 + \bar{u}^b\alpha_2\alpha_1\omega_1^{\dagger}\omega_1\\
&+& e^{-}\alpha_3\alpha_2\alpha_1\omega_1^{\dagger}\omega_1.
\end{eqnarray}

The $SU(3)$, written in terms of the ladder operators, are given by

\begin{eqnarray}
\nonumber\Lambda^{(1)}_1&=&-(\alpha_2^{\dagger}\alpha_1+\alpha_1^{\dagger}\alpha_2),\qquad\quad\;\; \Lambda^{(1)}_2=-i(\alpha_2^{\dagger}\alpha_1-\alpha_1^{\dagger}\alpha_2),\\
\nonumber\Lambda^{(1)}_3&=&\alpha_2^{\dagger}\alpha_2-\alpha_1^{\dagger}\alpha_1,\qquad\quad \Lambda^{(1)}_4=-(\alpha_1^{\dagger}\alpha_3+\alpha_3^{\dagger}\alpha_1),\\
\Lambda^{(1)}_5&=&i(\alpha_1^{\dagger}\alpha_3-\alpha_3^{\dagger}\alpha_1),\qquad \Lambda^{(1)}_6=-(\alpha_3^{\dagger}\alpha_2+\alpha_2^{\dagger}\alpha_3),\\
\nonumber\Lambda^{(1)}_7&=&-i(\alpha_3^{\dagger}\alpha_2-\alpha_2^{\dagger}\alpha_3),\qquad\;\;\;\; \Lambda^{(1)}_8=\frac{-1}{\sqrt{3}}(\alpha_1^{\dagger}\alpha_1+\alpha_2^{\dagger}\alpha_2-2\alpha_3^{\dagger}\alpha_3).
\end{eqnarray}
The $U(1)$ generator, written in terms of ladder operators is

\begin{eqnarray}
Q^{(1)}=\frac{1}{3}(\alpha_1^{\dagger}\alpha_1+\alpha_2^{\dagger}\alpha_2+\alpha_3^{\dagger}\alpha_3).
\end{eqnarray} 
%%%%%%%%%%%%%%%%%%%%%%%%%%%%%%%%%%
\subsection*{Generation 2}

\begin{eqnarray}
\nonumber\beta_1^{\dagger}&=&\frac{1}{2}(e_7+ie_8),\qquad \beta_2^{\dagger}=\frac{1}{2}(e_1+ie_{14}),\qquad \beta_3^{\dagger}=\frac{1}{2}(e_6+ie_9),\\
\nonumber\beta_1&=&\frac{1}{2}(-e_7+ie_8),\qquad \beta_2=\frac{1}{2}(-e_1+ie_{14}),\qquad \beta_3=\frac{1}{2}(-e_6+ie_9),\\
\nonumber \omega_2\omega_2^{\dagger}&=&\beta_1\beta_2\beta_3\beta_3^{\dagger}\beta_2^{\dagger}\beta_1^{\dagger}.
\end{eqnarray}

The second minimal left ideal is given by $S_2^u\equiv C\ell(6)_2\omega_2\omega_2^{\dagger}$. Explicitly,
\begin{eqnarray}
\nonumber S_2^u\equiv &{}&\\
\nonumber &{}&\;\;\nu_{\mu} \omega_2\omega_2^{\dagger}+\\
\nonumber \bar{s}^r\beta_1^{\dagger}\omega_2\omega_2^{\dagger} &+& \bar{s}^g\beta_2^{\dagger}\omega_2\omega_2^{\dagger} + \bar{s}^b\beta_3^{\dagger}\omega_2\omega_2^{\dagger}\\
\nonumber c^r\beta_3^{\dagger}\beta_2^{\dagger}\omega_2\omega_2^{\dagger} &+& c^g\beta_1^{\dagger}\beta_3^{\dagger}\omega_2\omega_2^{\dagger} + c^b\beta_2^{\dagger}\beta_1^{\dagger}\omega_2\omega_2^{\dagger}\\
&+& \mu^{+}\beta_3^{\dagger}\beta_2^{\dagger}\beta_1^{\dagger}\omega_2\omega_2^{\dagger},
\end{eqnarray}
The complex conjugate system gives a second linearly independent minimal left ideal
\begin{eqnarray}
\nonumber S_2^d\equiv &{}&\\
\nonumber &{}&\;\;\bar{\nu}_{\mu} \omega_2^{\dagger}\omega_2+\\
\nonumber s^r\beta_1\omega_2^{\dagger}\omega_2 &+& s^g\beta_2\omega_2^{\dagger}\omega_2 + s^b\beta_3\omega_2^{\dagger}\omega_2\\
\nonumber \bar{c}^r\beta_3\beta_2\omega_2^{\dagger}\omega_2 &+& \bar{c}^g\beta_1\beta_3\omega_2^{\dagger}\omega_2 + \bar{c}^b\beta_2\beta_1\omega_2^{\dagger}\omega_2\\
&+& \mu^{-}\beta_3\beta_2\beta_1\omega_2^{\dagger}\omega_2.
\end{eqnarray}

The $SU(3)$, written in terms of the ladder operators, are given by

\begin{eqnarray}
\nonumber\Lambda^{(2)}_1&=&-(\beta_2^{\dagger}\beta_1+\beta_1^{\dagger}\beta_2),\qquad\quad\;\; \Lambda^{(2)}_2=-i(\beta_2^{\dagger}\beta_1-\beta_1^{\dagger}\beta_2),\\
\nonumber\Lambda^{(2)}_3&=&\beta_2^{\dagger}\beta_2-\beta_1^{\dagger}\beta_1,\qquad\quad \Lambda^{(2)}_4=-(\beta_1^{\dagger}\beta_3+\beta_3^{\dagger}\beta_1),\\
\Lambda^{(2)}_5&=&i(\beta_1^{\dagger}\beta_3-\beta_3^{\dagger}\beta_1),\qquad \Lambda^{(2)}_6=-(\beta_3^{\dagger}\beta_2+\beta_2^{\dagger}\beta_3),\\
\nonumber\Lambda^{(2)}_7&=&-i(\beta_3^{\dagger}\beta_2-\beta_2^{\dagger}\beta_3),\qquad\;\;\;\; \Lambda^{(2)}_8=\frac{-1}{\sqrt{3}}(\beta_1^{\dagger}\beta_1+\beta_2^{\dagger}\beta_2-2\beta_3^{\dagger}\beta_3).
\end{eqnarray}

The $U(1)$ generator, written in terms of ladder operators is

\begin{eqnarray}
Q^{(2)}=\frac{1}{3}(\beta_1^{\dagger}\beta_1+\beta_2^{\dagger}\beta_2+\beta_3^{\dagger}\beta_3).
\end{eqnarray} 

%%%%%%%%%%%%%%%%%%%%%%%%%%%%%%%%%%
\subsection*{Generation 3}

\begin{eqnarray}
\nonumber\gamma_1^{\dagger}&=&\frac{1}{2}(e_{13}+ie_2),\qquad \gamma_2^{\dagger}=\frac{1}{2}(e_1+ie_{14}),\qquad \gamma_3^{\dagger}=\frac{1}{2}(e_{12}+ie_3),\\
\nonumber\gamma_1&=&\frac{1}{2}(-e_{13}+ie_2),\qquad \gamma_2=\frac{1}{2}(-e_1+ie_{14}),\qquad \gamma_3=\frac{1}{2}(-e_{12}+ie_3),\\
\nonumber \omega_3\omega_3^{\dagger}&=&\gamma_1\gamma_2\gamma_3\gamma_3^{\dagger}\gamma_2^{\dagger}\gamma_1^{\dagger}.
\end{eqnarray}

The third minimal left ideal is given by $S_3^u\equiv C\ell(6)_3\omega_3\omega_3^{\dagger}$. Explicitly,
\begin{eqnarray}
\nonumber S_3^u\equiv &{}&\\
\nonumber &{}&\;\;\nu_{\tau} \omega_3\omega_3^{\dagger}+\\
\nonumber \bar{b}^r\gamma_1^{\dagger}\omega_3\omega_3^{\dagger} &+& \bar{b}^g\gamma_2^{\dagger}\omega_3\omega_3^{\dagger} + \bar{b}^b\gamma_3^{\dagger}\omega_3\omega_3^{\dagger}\\
\nonumber t^r\gamma_3^{\dagger}\gamma_2^{\dagger}\omega_3\omega_3^{\dagger} &+& t^g\gamma_1^{\dagger}\gamma_3^{\dagger}\omega_3\omega_3^{\dagger} + t^b\gamma_2^{\dagger}\gamma_1^{\dagger}\omega_3\omega_3^{\dagger}\\
&+& \tau^{+}\gamma_3^{\dagger}\gamma_2^{\dagger}\gamma_1^{\dagger}\omega_3\omega_3^{\dagger},
\end{eqnarray}
The complex conjugate system gives a second linearly independent minimal left ideal
\begin{eqnarray}
\nonumber S_3^d\equiv &{}&\\
\nonumber &{}&\;\;\bar{\nu}_{\tau} \omega_3^{\dagger}\omega_3+\\
\nonumber b^r\gamma_1\omega_3^{\dagger}\omega_3 &+& b^g\gamma_2\omega_3^{\dagger}\omega_3 + b^b\gamma_3\omega_3^{\dagger}\omega_3\\
\nonumber \bar{t}^r\gamma_3\gamma_2\omega_3^{\dagger}\omega_3 &+& \bar{t}^g\gamma_1\gamma_3\omega_3^{\dagger}\omega_3 + \bar{t}^b\gamma_2\gamma_1\omega_3^{\dagger}\omega_3\\
&+& \tau^{-}\gamma_3\gamma_2\gamma_1\omega_3^{\dagger}\omega_3.
\end{eqnarray}

The $SU(3)$, written in terms of the ladder operators, are given by

\begin{eqnarray}
\nonumber\Lambda^{(3)}_1&=&-(\gamma_2^{\dagger}\gamma_1+\gamma_1^{\dagger}\gamma_2),\qquad\quad\;\; \Lambda^{(3)}_2=-i(\gamma_2^{\dagger}\gamma_1-\gamma_1^{\dagger}\gamma_2),\\
\nonumber\Lambda^{(3)}_3&=&\gamma_2^{\dagger}\gamma_2-\gamma_1^{\dagger}\gamma_1,\qquad\quad \Lambda^{(3)}_4=-(\gamma_1^{\dagger}\gamma_3+\gamma_3^{\dagger}\gamma_1),\\
\Lambda^{(3)}_5&=&i(\gamma_1^{\dagger}\gamma_3-\gamma_3^{\dagger}\gamma_1),\qquad \Lambda^{(3)}_6=-(\gamma_3^{\dagger}\gamma_2+\gamma_2^{\dagger}\gamma_3),\\
\nonumber\Lambda^{(3)}_7&=&-i(\gamma_3^{\dagger}\gamma_2-\gamma_2^{\dagger}\gamma_3),\qquad\;\;\;\; \Lambda^{(3)}_8=\frac{-1}{\sqrt{3}}(\gamma_1^{\dagger}\gamma_1+\gamma_2^{\dagger}\gamma_2- 2\gamma_3^{\dagger}\gamma_3).
\end{eqnarray}

The $U(1)$ generator, written in terms of ladder operators is

\begin{eqnarray}
Q^{(3)}=\frac{1}{3}(\gamma_1^{\dagger}\gamma_1+\gamma_2^{\dagger}\gamma_2+\gamma_3^{\dagger}\gamma_3).
\end{eqnarray} 

%\section*{References}
%%%%%%%%%%%%%%%%%%%%%%%%%%%%%%%%%%%%%%%%%%%%%%%%%%%%%%%%%%%%%%%%%%%%%
\bibliography{Jordanalgebra}  % Replace xxx by your  usercode (no extension)

\begin{thebibliography}{10}

\bibitem{hurwitz1922komposition}
Adolf Hurwitz.
\newblock {\"U}ber die komposition der quadratischen formen.
\newblock {\em Mathematische Annalen}, 88(1-2):1--25, 1922.

\bibitem{zorn1931theorie}
Max Zorn.
\newblock Theorie der alternativen ringe.
\newblock In {\em Abhandlungen aus dem Mathematischen Seminar der
  Universit{\"a}t Hamburg}, volume~8, pages 123--147. Springer, 1931.

\bibitem{gunaydin1973quark}
Murat G{\"u}naydin and Feza G{\"u}rsey.
\newblock Quark structure and octonions.
\newblock {\em Journal of Mathematical Physics}, 14(11):1651--1667, 1973.

\bibitem{gunaydin1974quark}
M~G{\"u}naydin and F~G{\"u}rsey.
\newblock Quark statistics and octonions.
\newblock {\em Physical Review D}, 9(12):3387, 1974.

\bibitem{baez2002octonions}
John Baez.
\newblock The octonions.
\newblock {\em Bulletin of the American Mathematical Society}, 39(2):145--205,
  2002.

\bibitem{duff1998octonionic}
MJ~Duff, Jonathan~M Evans, Ramzi~R Khuri, JX~Lu, and Ruben Minasian.
\newblock The octonionic membrane.
\newblock {\em Nuclear Physics B-Proceedings Supplements}, 68(1-3):295--302,
  1998.

\bibitem{baez2009division}
John~C Baez and John Huerta.
\newblock Division algebras and supersymmetry i.
\newblock {\em Superstrings, geometry, topology, and C*-algebras}, 81:65--80,
  2009.

\bibitem{baez2011division}
John~C Baez, John Huerta, et~al.
\newblock Division algebras and supersymmetry ii.
\newblock {\em Advances in Theoretical and Mathematical Physics},
  15(5):1373--1410, 2011.

\bibitem{anastasiou2014super}
Alexandros Anastasiou, Leron Borsten, MJ~Duff, LJ~Hughes, and Silvia Nagy.
\newblock Super yang-mills, division algebras and triality.
\newblock {\em Journal of High Energy Physics}, 2014(8):80, 2014.

\bibitem{dixon2013division}
Geoffrey~M Dixon.
\newblock {\em Division Algebras:: Octonions Quaternions Complex Numbers and
  the Algebraic Design of Physics}, volume 290.
\newblock Springer Science \& Business Media, 2013.

\bibitem{furey2016standard}
Cohl Furey.
\newblock Standard model physics from an algebra?
\newblock {\em arXiv preprint arXiv:1611.09182}, 2016.

\bibitem{gresnigt2018braids}
Niels~G Gresnigt.
\newblock Braids, normed division algebras, and standard model symmetries.
\newblock {\em Phys. Lett. B, 783, 212-221 (arXiv preprint arXiv:1803.02202)},
  2018.

\bibitem{gresnigt2018braided}
Niels~G Gresnigt.
\newblock Braided fermions from hurwitz algebras.
\newblock {\em arXiv preprint arXiv:1901.01312}, 2018.

\bibitem{dray2010octonionic}
Tevian Dray and Corinne~A Manogue.
\newblock Octonionic cayley spinors and e6.
\newblock {\em Comment. Math. Univ. Carolin}, 51(2):193--207, 2010.

\bibitem{burdik2017revisiting}
C~Burdik, S~Catto, Y~G{\"u}rcan, A~Khalfan, and L~Kurt.
\newblock Revisiting the role of octonions in hadronic physics.
\newblock {\em Physics of Particles and Nuclei Letters}, 14(2):390--394, 2017.

\bibitem{burdik2018hurwitz}
{\v{C}}estmir Burdik and Sultan Catto.
\newblock Hurwitz algebras and the octonion algebra.
\newblock In {\em Journal of Physics: Conference Series}, volume 965, page
  012009. IOP Publishing, 2018.

\bibitem{catto2018quantum}
Sultan Catto, Yasemin G{\"u}rcan, Amish Khalfan, and Levent Kurt.
\newblock Quantum symmetries: From clifford and hurwitz algebras to m-theory
  and leech lattices.
\newblock {\em Advances in Applied Clifford Algebras}, 28(4):81, 2018.

\bibitem{ablamowicz1995construction}
Rafa{\l} Ab{\l}amowicz.
\newblock Construction of spinors via witt decomposition and primitive
  idempotents: a review.
\newblock In {\em Clifford Algebras and Spinor Structures}, pages 113--123.
  Springer, 1995.

\bibitem{Stoica2018}
Ovidiu~Cristinel Stoica.
\newblock Leptons, quarks, and gauge from the complex clifford algebra cl(6).
\newblock {\em Advances in Applied Clifford Algebras}, 28(3):52, 2018.

\bibitem{furey2014generations}
Cohl Furey.
\newblock Generations: three prints, in colour.
\newblock {\em Journal of High Energy Physics}, 2014(10):1--11, 2014.

\bibitem{furey2018three}
C~Furey.
\newblock Three generations, two unbroken gauge symmetries, and one
  eight-dimensional algebra.
\newblock {\em Physics Letters B}, 785:84--89, 2018.

\bibitem{dubois2016exceptional}
Michel Dubois-Violette.
\newblock Exceptional quantum geometry and particle physics.
\newblock {\em Nuclear Physics B}, 912:426--449, 2016.

\bibitem{todorov2018deducing}
Ivan Todorov and Michel Dubois-Violette.
\newblock Deducing the symmetry of the standard model from the automorphism and
  structure groups of the exceptional jordan algebra.
\newblock {\em International Journal of Modern Physics A}, 33(20):1850118,
  2018.

\bibitem{dubois2019exceptional}
Michel Dubois-Violette and Ivan Todorov.
\newblock Exceptional quantum geometry and particle physics ii.
\newblock {\em Nuclear Physics B}, 938:751--761, 2019.

\bibitem{dixon2010division}
Geoffrey~M Dixon.
\newblock Division algebras; spinors; idempotents; the algebraic structure of
  reality.
\newblock {\em arXiv preprint arXiv:1012.1304}, 2010.

\bibitem{dixon1990derivation}
Geoffrey Dixon.
\newblock Derivation of the standard model.
\newblock {\em Il Nuovo Cimento B (1971-1996)}, 105(3):349--364, 1990.

\bibitem{dixon2004division}
Geoffrey Dixon.
\newblock Division algebras: Family replication.
\newblock {\em Journal of mathematical physics}, 45(10):3878--3882, 2004.

\bibitem{manogue1999dimensional}
Corinne~A Manogue and Tevian Dray.
\newblock Dimensional reduction.
\newblock {\em Modern Physics Letters A}, 14(02):99--103, 1999.

\bibitem{dray2000quaternionic}
Tevian Dray and Corinne~A Manogue.
\newblock Quaternionic spin.
\newblock In {\em Clifford Algebras and their Applications in Mathematical
  Physics}, pages 21--37. Springer, 2000.

\bibitem{moreno1997zero}
R~Guillermo Moreno.
\newblock The zero divisors of the cayley-dickson algebras over the real
  numbers.
\newblock {\em arXiv preprint q-alg/9710013}, 1997.

\bibitem{dixon2017division}
Geoffrey~M Dixon.
\newblock {\em Division algebras, lattices, physics, windmill tilting}.
\newblock Dixon, 2017.

\bibitem{lounesto2001caa}
P.~Lounesto.
\newblock {\em {Clifford Algebras and Spinors}}.
\newblock Cambridge University Press, 2001.

\bibitem{manogue2010octonions}
Corinne~A Manogue and Tevian Dray.
\newblock Octonions, e6, and particle physics.
\newblock In {\em Journal of Physics: Conference Series}, volume 254, page
  012005. IOP Publishing, 2010.

\end{thebibliography}
\bibliographystyle{unsrt}  
%%%%%%%%%%%%%%%%%%%%%%%%%%%%%%%%%%%%%%%%%%%%%%%%%%%%%%%%%%%%%%%%%%%%%

\end{document}